\documentstyle[twoside]{article}

\catcode`\@=11
\long\def\@makefntext#1{
\protect\noindent \hbox to 3.2pt {\hskip-.9pt  
$^{{\eightrm\@thefnmark}}$\hfil}#1\hfill}		

\def\@makefnmark{\hbox to 0pt{$^{\@thefnmark}$\hss}}	
	
\def\ps@myheadings{\let\@mkboth\@gobbletwo
\def\@oddhead{\hbox{}
\rightmark\hfil\eightrm\thepage}   
\def\@oddfoot{}\def\@evenhead{\eightrm\thepage\hfil
\leftmark\hbox{}}\def\@evenfoot{}
\def\sectionmark##1{}\def\subsectionmark##1{}}

\oddsidemargin=\evensidemargin
\newcounter{sectionc}\newcounter{subsectionc}\newcounter{subsubsectionc}
\renewcommand{\section}[1] {\vspace{12pt}\addtocounter{sectionc}{1} 
\setcounter{subsectionc}{0}\setcounter{subsubsectionc}{0}\noindent 
	{\tenbf\thesectionc. #1}\par\vspace{5pt}}
\renewcommand{\subsection}[1] {\vspace{12pt}\addtocounter{subsectionc}{1} 
	\setcounter{subsubsectionc}{0}\noindent 
	{\bf\thesectionc.\thesubsectionc. {\kern1pt \bfit #1}}\par\vspace{5pt}}
\renewcommand{\subsubsection}[1] {\vspace{12pt}\addtocounter{subsubsectionc}{1}
	\noindent{\tenrm\thesectionc.\thesubsectionc.\thesubsubsectionc.
	{\kern1pt \tenit #1}}\par\vspace{5pt}}
\newcommand{\nonumsection}[1] {\vspace{12pt}\noindent{\tenbf #1}
	\par\vspace{5pt}}

\newcounter{appendixc}
\newcounter{subappendixc}[appendixc]
\newcounter{subsubappendixc}[subappendixc]
\renewcommand{\thesubappendixc}{\Alph{appendixc}.\arabic{subappendixc}}
\renewcommand{\thesubsubappendixc}
	{\Alph{appendixc}.\arabic{subappendixc}.\arabic{subsubappendixc}}

\renewcommand{\appendix}[1] {\vspace{12pt}
        \refstepcounter{appendixc}
        \setcounter{figure}{0}
        \setcounter{table}{0}
        \setcounter{lemma}{0}
        \setcounter{theorem}{0}
        \setcounter{corollary}{0}
        \setcounter{definition}{0}
        \setcounter{equation}{0}
        \renewcommand{\thefigure}{\Alph{appendixc}.\arabic{figure}}
        \renewcommand{\thetable}{\Alph{appendixc}.\arabic{table}}
        \renewcommand{\theappendixc}{\Alph{appendixc}}
        \renewcommand{\thelemma}{\Alph{appendixc}.\arabic{lemma}}
        \renewcommand{\thetheorem}{\Alph{appendixc}.\arabic{theorem}}
        \renewcommand{\thedefinition}{\Alph{appendixc}.\arabic{definition}}
        \renewcommand{\thecorollary}{\Alph{appendixc}.\arabic{corollary}}
        \renewcommand{\theequation}{\Alph{appendixc}.\arabic{equation}}
        \noindent{\tenbf Appendix \theappendixc #1}\par\vspace{5pt}}
\newcommand{\subappendix}[1] {\vspace{12pt}
        \refstepcounter{subappendixc}
        \noindent{\bf Appendix \thesubappendixc. {\kern1pt \bfit #1}}
	\par\vspace{5pt}}
\newcommand{\subsubappendix}[1] {\vspace{12pt}
        \refstepcounter{subsubappendixc}
        \noindent{\rm Appendix \thesubsubappendixc. {\kern1pt \tenit #1}}
	\par\vspace{5pt}}

\newcommand{\textlineskip}{\baselineskip=13pt}
\newcommand{\smalllineskip}{\baselineskip=10pt}

\def\eightcirc{
\begin{picture}(0,0)
\put(4.4,1.8){\circle{6.5}}
\end{picture}}
\def\eightcopyright{\eightcirc\kern2.7pt\hbox{\eightrm c}}

\def\abstracts#1#2#3{{
	\centering{\begin{minipage}{4.5in}\baselineskip=10pt\footnotesize
	\parindent=0pt #1\par 
	\parindent=15pt #2\par
	\parindent=15pt #3
	\end{minipage}}\par}} 

\newcommand{\bibit}{\nineit}

\renewenvironment{thebibliography}[1]
	{\frenchspacing
	 \ninerm\baselineskip=11pt
	 \begin{list}{\arabic{enumi}.}
	{\usecounter{enumi}\setlength{\parsep}{0pt}
	 \setlength{\leftmargin 12.7pt}{\rightmargin 0pt} 
	 \setlength{\itemsep}{0pt} \settowidth
	{\labelwidth}{#1.}\sloppy}}{\end{list}}

\newcounter{itemlistc}
\newcounter{romanlistc}
\newcounter{alphlistc}
\newcounter{arabiclistc}

\newcommand{\fcaption}[1]{
        \refstepcounter{figure}
        \setbox\@tempboxa = \hbox{\footnotesize Fig.~\thefigure. #1}
        \ifdim \wd\@tempboxa > 5in
           {\begin{center}
        \parbox{5in}{\footnotesize\smalllineskip Fig.~\thefigure. #1}
            \end{center}}
        \else
             {\begin{center}
             {\footnotesize Fig.~\thefigure. #1}
              \end{center}}
        \fi}

\newcommand{\tcaption}[1]{
        \refstepcounter{table}
        \setbox\@tempboxa = \hbox{\footnotesize Table~\thetable. #1}
        \ifdim \wd\@tempboxa > 5in
           {\begin{center}
        \parbox{5in}{\footnotesize\smalllineskip Table~\thetable. #1}
            \end{center}}
        \else
             {\begin{center}
             {\footnotesize Table~\thetable. #1}
              \end{center}}
        \fi}

\def\@citex[#1]#2{\if@filesw\immediate\write\@auxout
	{\string\citation{#2}}\fi
\def\@citea{}\@cite{\@for\@citeb:=#2\do
	{\@citea\def\@citea{,}\@ifundefined
	{b@\@citeb}{{\bf ?}\@warning
	{Citation `\@citeb' on page \thepage \space undefined}}
	{\csname b@\@citeb\endcsname}}}{#1}}

\newif\if@cghi
\def\cite{\@cghitrue\@ifnextchar [{\@tempswatrue
	\@citex}{\@tempswafalse\@citex[]}}
\def\citelow{\@cghifalse\@ifnextchar [{\@tempswatrue
	\@citex}{\@tempswafalse\@citex[]}}
\def\@cite#1#2{{$\null^{#1}$\if@tempswa\typeout
	{IJCGA warning: optional citation argument 
	ignored: `#2'} \fi}}

\def\pmb#1{\setbox0=\hbox{#1}
	\kern-.025em\copy0\kern-\wd0
	\kern.05em\copy0\kern-\wd0
	\kern-.025em\raise.0433em\box0}

\def\fnt#1#2{\footnotetext{\kern-.3em
	{$^{\mbox{\scriptsize #1}}$}{#2}}}

\def\fpage#1{\begingroup
\voffset=.3in
\thispagestyle{empty}\begin{table}[b]\centerline{\footnotesize #1}
	\end{table}\endgroup}

\def\runninghead#1#2{\pagestyle{myheadings}
\markboth{{\protect\footnotesize\it{\quad #1}}\hfill}
{\hfill{\protect\footnotesize\it{#2\quad}}}}
\font\tenrm=cmr10
\font\tenit=cmti10 
\font\tenbf=cmbx10
\font\bfit=cmbxti10 at 10pt
\font\ninerm=cmr9
\font\nineit=cmti9

\font\eightrm=cmr8

\def\qed{\hbox{${\vcenter{\vbox{			
   \hrule height 0.4pt\hbox{\vrule width 0.4pt height 6pt
   \kern5pt\vrule width 0.4pt}\hrule height 0.4pt}}}$}}

\begin{document}

\catcode`@=11
\def\@cite#1{${}^{\mbox{\small #1}}$}
\catcode`@=11

\runninghead{QUARKONIUM POLARIZATION $\ldots$}
	    {QUARKONIUM POLARIZATION $\ldots$}

\normalsize\textlineskip
\thispagestyle{empty}
\setcounter{page}{1}

\hfill \vbox{\halign{&#\hfil\cr
		& OHSTPY-HEP-T-96-025	\cr
		& September 1996	\cr}}
\vspace*{0.44truein}

\fpage{1}
\centerline{\bf QUARKONIUM POLARIZATION }
\vspace*{0.035truein}
\centerline{\bf IN THE NRQCD FACTORIZATION FRAMEWORK\footnote{Invited talk
	presented at the Quarkonium Physics Workshop, 
	University of Illinois at Chicago, June 1996.}}
\vspace*{0.37truein}
\centerline{\footnotesize ERIC BRAATEN}
\vspace*{0.015truein}
\centerline{\footnotesize\it Department of Physics, Ohio State University}
\baselineskip=10pt
\centerline{\footnotesize\it Columbus, OH 43210, USA}

\vspace*{0.21truein}
\abstracts{The NRQCD factorization approach for calculating inclusive 
production of heavy quarkonium gives unambiguous 
predictions for the polarization of quarkonium states.
The factorization formula for polarized states can be obtained 
by using the threshold expansion method to calculate the 
short-distance coefficients and then using symmetries of NRQD 
to reduce the NRQCD matrix elements.  A particularly dramatic
prediction of the NRQCD factorization framework is that prompt 
psi's and psi-primes's produced at the Tevatron should be 
predominantly transversely polarized at large transverse momentum.
}{}{}

\vspace*{1pt}\textlineskip	
\section{Introduction}
\noindent
The NRQCD factorization approach\cite{B-B-L} is a systematic framework
for analyzing the inclusive cross sections and annihilation decay rates 
of heavy quarkonium.  The cross section (or decay rate) is factored 
into short-distance coefficients that are computable using perturbation 
theory and long-distance NRQCD matrix elements.  The matrix elements 
scale as a definite power of $v$, the typical relative velocity of the 
heavy quark in quarkonium.  Thus the cross section can be organized
into a double expansion in powers of $v$ and powers of $\alpha_s(m_c)$, 
where $m_c$ is the heavy quark mass.  

The NRQCD factorization framework makes 
definite, and in some cases rather dramatic, 
predictions for the dependence of the cross section on 
the polarization of the quarkonium.  Below, I outline the 
NRQCD factorization approach as it applies to polarized quarkonium states
and then summarize the applications that have been carried out thus far. 

\section{NRQCD Factorization}
\noindent
The general expression for the cross section for the inclusive production
of a quarkonium state $H$ with four-momentum $P$ is
\begin{eqnarray}
\sum_X d \sigma(12 \to H(P) + X) &=&
{1 \over 4 E_1 E_2 v_{12}}\;  {d^3P \over (2 \pi)^3 2 E_P}
\nonumber \\
&& \hspace{-1in} \times 
\sum_X \; (2 \pi)^4 \delta^4(k_1 + k_2 - P - k_X) \;
	|{\cal T}_{1 2 \to H(P) + X}|^2 \,, 
\label{dsig}
\end{eqnarray}
where ${\cal T}_{1 2 \to H(P) + X}$ is the T-matrix element
and the sum on the right side includes integration over the 
phase space of the additional particles.
This cross section involves  both ``short distances'' of order 
$1/m_c$ or smaller and ``long distances''  of order $1/(m_c v)$ 
or larger.  
The production of the $c \bar c$ pair involves short distances,
because the  parton processes that produce 
a $c \bar c$ pair always involve particles that are off their mass shells
by amounts of order $m_c$ and can therefore propagate only over
short distances.  
The binding of the $c$ and $\bar c$ into the state $H$
involves long distances, because gluons whose wavelengths are comparable to 
or larger than the size of the bound state, which is of 
order $1/(m_c v)$, play a large role in the binding. 

If the cross section in (\ref{dsig}) is sufficiently inclusive, 
short-distance and long-distance effects can be factored\cite{B-B-L}.
(If there are hadrons in the initial state, it is also necessary
to restrict the four-momentum of the quarkonium to be significantly 
different from that of the initial hadrons in order to avoid 
contributions from diffractive scattering.) 
Like many other factorization ``theorems'' of perturbative QCD,
the factorization of quarkonium cross sections has not been 
proven with complete rigor, but it is a plausible generalization 
of the factorization theorems for the Drell-Yan production of 
muon pairs and for heavy quark production.  
The factorization relies on cancellations between soft
partons that are emitted by the $c \bar c$ pair and 
soft partons that are exchanged between the $c \bar c$ pair 
and other jet-like collections of collinear partons.
After taking into account these cancellations,
the cross section can be factored into short-distance
and long-distance parts.  The short-distance
parts involve the production of a $c \bar c$ pair with small 
relative momentum plus hard partons.  The $c \bar c$ pair
that emerges from the short-distance part  
is essentially pointlike on the scale of the quarkonium wavefunction.
The long-distance parts involve the formation of
$H$ plus soft partons from the pointlike $c \bar c$ pair. 
  
The standard factorization methods of perturbative QCD produce an
expression for the cross section that involves an integral
over the relative momentum of the $c$ and $\bar c$ that form 
the quarkonium.  
Long-distance and short-distance effects can be further untangled by 
expanding the short-distance factors in powers of the relative momentum 
${\bf q}$ and absorbing the  integration over ${\bf q}$ into the 
long-distance parts.  The resulting long-distance
factors can be expressed as matrix elements in an effective 
field theory called {\it nonrelativistic QCD} (NRQCD).  
The T-matrix elements in (\ref{dsig}) that survive after 
soft-parton cancellations can be expressed in the form
\begin{equation}
{\cal T}_{1 2 \to H(P) + X_H + X_S}  \;\approx\;
\sum_n \hat{\cal T}_{1 2 \to c \bar c(P,n) + X_H} 
\langle H + X_S | \psi^\dagger {\cal K}_n \chi (x=0) | 0 \rangle \,, 
\label{T-fact}
\end{equation}
where the sum includes all possible color and angular-momentum states of the 
$c \bar c$ pair.  The factor $\hat{\cal T}_{1 2 \to c \bar c(P,n) + X_H}$
can be interpreted as a T-matrix element for producing a $c \bar c$ pair
in the state $n$ plus the hard partons $X_H$.  The second factor on the right
is a matrix element in NRQCD between the vacuum state 
and a state that in the asymptotic future consists of the 
quarkonium $H$ at rest plus the soft partons $X_S$.  The local operator
$\psi^\dagger {\cal K}_n \chi$ creates a pointlike $c \bar c$ pair 
in the state $n$. 

After inserting the expression (\ref{T-fact}) for the T-matrix elements
into (\ref{dsig}),  we can rearrange the cross section into a form 
in which the short-distance and long-distance contributions are factored:
\begin{eqnarray}
\sum_X \; (2 \pi)^4 \delta^4(k_1 + k_2 - P - k_X) \;
	| {\cal T}_{1 2 \to H(P) + X}|^2 && 
\nonumber \\
&& \hspace{-2.5in} \;\approx\; \sum_{m n}
\left( \sum_{X_H} 
	(\hat{\cal T}_{1 2 \to c \bar c(P,m) + X_H})^* 
	\hat{\cal T}_{1 2 \to c \bar c(P,n) + X_H} \right)
\nonumber \\
&& \hspace{-2in} \times
\left( \sum_{X_S}
	\langle 0 | \chi^\dagger {\cal K}^\dagger_m \psi | H + X_S \rangle 
	\langle H + X_S | \psi^\dagger {\cal K}_n \chi | 0 \rangle 
\right) \,. 
\label{TT-fact}
\end{eqnarray}
The short-distance factor is
\begin{equation}
C_{mn}(k_1,k_2, P) \;=\; \sum_{X_H} 
	(\hat{\cal T}_{1 2 \to c \bar c(P,m) + X_H})^* 
	\hat{\cal T}_{1 2 \to c \bar c(P,n) + X_H}  \,. 
\label{Cmn}
\end{equation}
The long-distance factor is
\begin{equation}
\langle {\cal O}^H_{mn} \rangle \;=\; 
\langle 0 | \chi^\dagger {\cal K}^\dagger_m \psi \; {\cal P}_H \;
	 \psi^\dagger {\cal K}_n \chi | 0 \rangle \,, 
\label{O-H}
\end{equation}
where ${\cal P}_H$ projects onto states that in the asymptotic future
contain the quarkonium state $H$ plus soft partons:
\begin{equation}
{\cal P}_H \;=\;
\sum_{X_S} | H + X_S \rangle \langle H + X_S |  \,. 
\label{P-H}
\end{equation}
Inserting (\ref{TT-fact}) into 
(\ref{dsig}), we obtain the NRQCD factorization formula for the inclusive
cross section:
\begin{equation}
\sum_X d \sigma(12 \to H(P) + X) \;=\;
{1 \over 4 E_1 E_2 v_{12}}\;  {d^3P \over (2 \pi)^3 2 E_P}
\sum_{mn}  C_{mn}(k_1,k_2, P) \; \langle {\cal O}^H_{mn} \rangle \,. 
\label{dsig-fact}
\end{equation}

We can write a similar factorization formula for the inclusive 
decay rate of the quarkonium $H$ via the annihilation 
of the $c \bar c$ pair.  The general formula for the decay rate is
\begin{equation}
\sum_X d \Gamma(H \to X) \;=\;
{1 \over 2 M_H} \;  
\sum_X \; (2 \pi)^4 \delta^4(P - k_X) \;
	|{\cal T}_{H(P) \to X}|^2 \,, 
\label{dGam}
\end{equation}
where $P = (M_H,{\bf 0})$.  This can be expressed in the factored form
\begin{equation}
\sum_X d \Gamma(H \to X) \;=\;
{1 \over 2 M_H}
\sum_{mn}  C_{mn} \; \langle H | {\cal O}_{mn} | H \rangle \,, 
\label{dGam-fact}
\end{equation}
where the $C_{mn}$'s are short-distance coefficients.
The long-distance factors 
are expectation values in the quarkonium state
of local four-fermion operators of the form
${\cal O}_{mn} = \psi^\dagger {\cal K}^\dagger_m \chi 
	 \chi^\dagger {\cal K}_n \psi$.

\section{Short-distance Coefficients}
\noindent
Since the coefficients $C_{mn}$ in (\ref{dsig-fact}) and (\ref{dGam-fact})
involve only short distances of order $1/m_c$ or larger, they can
be expressed as perturbation series in $\alpha_s(m_c)$.
These coefficients are known at tree level for many processes,
and in a few cases they are known at the one-loop level.
Most of these coefficients have been obtained by calculating
the perturbative cross section for producing a $c \bar c$ pair in a state
with a prescribed nonrelativistic wavefunction.  
The inclusive cross section is sensitive only to the 
behavior of the wavefunction near the origin.
The analogs of the NRQCD matrix elements (\ref{O-H})
for the $c \bar c$  state can also be calculated using 
perturbation theory in terms of the  prescribed wavefunction.  
Knowing the cross section and the matrix elements,
we can read off the short-distance coefficients. 
Unfortunately, this method is not sufficiently general to
determine all the short-distance coefficients, especially for
polarized quarkonium states.

The {\it threshold expansion method}, developed recently
by Braaten and Chen\cite{Braaten-Chen}, is a general 
prescription for calculating the short-distance coefficients.
It is based directly on the NRQCD factorization approach, and   
can be readily applied to polarized quarkonium.

\subsection{Threshold expansion method}
\noindent
The threshold expansion method relies on the fact that the 
short-distance coefficients in (\ref{TT-fact}) are insensitive to
the long-distance effects that bind the $c \bar c$ pairs into the 
quarkonium state $H$.  Thus the factorization formula will hold 
with the same short-distance coefficients if we replace $H$
by asymptotic perturbative states 
$c \bar c = c \bar c({\bf q},\xi,\eta)$ that consist 
of a $c$ and a $\bar c$ with relative momentum ${\bf q}$ and spin/color
state that is represented by the Pauli spinors $\xi$ and $\eta$.
To completely determine the short-distance coefficients, we need to use 
different states $c \bar c$ and $c \bar c'$ in the T-matrix element 
and in its complex conjugate.  The resulting matching prescription is 
\begin{eqnarray}
\sum_X \; (2 \pi)^4 \delta^4(k_1 + k_2 - P - k_X) \;
	({\cal T}_{1 2 \to c \bar c'(P) + X})^* 
	{\cal T}_{1 2 \to  c \bar c(P) + X} \Big|_{pQCD}  && 
\nonumber \\
&& \hspace{-3.5in} \;\approx\; \sum_{m n}
C_{mn}(k_1,k_2,P)
	\langle 0 | \chi^\dagger {\cal K}_m \psi  \; 
	{\cal P}_{c \bar c',c \bar c} \;
	\psi^\dagger {\cal K}_n \chi | 0 \rangle \Big|_{pNRQCD} \,, 
\label{TT-match}
\end{eqnarray}
where the projection operator in the matrix element is 
\begin{equation}
{\cal P}_{c \bar c',c \bar c} \;=\;
\sum_{X_S} |  c \bar c' + X_S \rangle 
		\langle c \bar c + X_S |  \,. 
\label{P-ccbar}
\end{equation}
The left side of (\ref{TT-match}) is to be calculated using perturbative
QCD, and then expanded in powers of the relative momenta
${\bf q}$ and ${\bf q}'$.
The matrix elements on the right side are to be calculated using 
perturbative NRQCD, and then expanded in powers of ${\bf q}$ and ${\bf q}'$.
The coefficients $C_{mn}$ are then determined by matching these
expansions order by order in $\alpha_s$.

\subsection{Example}
\noindent
We illustrate the threshold expansion method by carrying out one of the 
simplest matching calculations.  It gives the 
short-distance coefficient corresponding to 
the parton process $q \bar q \to c \bar c$.  The  T-matrix element
for this process is 
\begin{equation}
{\cal T}_{1 2 \to c \bar c} 
\;=\; g^2 {1 \over P^2} \; {\bar v}(k_2) \gamma_\mu T^a u(k_1) \;
	\bar u(p) \gamma^\mu T^a v({\bar p}) .
\label{T-qqbar}
\end{equation}
Making a nonrelativistic expansion of the spinors of the $c$ and $\bar c$,
this reduces to 
\begin{equation}
{\cal T}_{1 2 \to c \bar c} 
\;=\; {g^2  \over 2 m_c} \; {\bar v}(k_2) \gamma_\mu T^a u(k_1) \;
	L^\mu_{\ i} \; \xi^\dagger \sigma^i T^a \eta ,
\label{T-qqbar0}
\end{equation}
where $L^\mu_{\ i}$ are elements of the boost matrix that transforms from
the rest frame of the $c \bar c$ pair to the frame in which it has total 
four-momentum $P$.  Multiplying by the complex conjugate of 
${\cal T}_{1 2 \to c \bar c'}$ and averaging over initial spins and
colors, we obtain 
\begin{equation}
({\cal T}_{1 2 \to c \bar c'})^* \; 
	{\cal T}_{1 2 \to c \bar c}
\;=\;  {4 \pi^2 \alpha_s^2 \over 9} 
	\left[ \delta^{ji} - {\hat z}^j {\hat z}^i \right]
	{\eta'}^\dagger \sigma^j T^a \xi' \xi^\dagger \sigma^i T^a \eta ,	
\label{TT-qqbar}
\end{equation}
where $\bf {\hat z}$ is a unit vector in the direction of the momenta
of the colliding $c$ and $\bar c$.  The spinor factor can be expressed 
in terms of an NRQCD matrix element:
\begin{equation}
\langle \chi^\dagger \sigma^j T^a \psi 
	| c \bar c' \rangle \langle c \bar c | 
	\psi^\dagger \sigma^i T^a \chi \rangle  \Big|_{pNRQCD}
\;=\; 4 m_c^2 \;  {\eta'}^\dagger \sigma^j T^a \xi' 
	\xi^\dagger \sigma^i T^a \eta \,.
\label{M-qqbar}
\end{equation}
Using the matching prescription (\ref{TT-match}),
the  short-distance coefficient of the matrix element 
$\langle \chi^\dagger \sigma^j T^a \psi {\cal P}_{c \bar c',c \bar c} 
	\psi^\dagger \sigma^i T^a \chi \rangle$
is
\begin{equation}
C_{ij} \;=\;  
(2 \pi)^4 \delta^4(k_1 + k_2 - P) \; 
{ \pi^2 \alpha_s^2 \over 9 m_c^2} 
	\left[ \delta^{ji} - {\hat z}^j {\hat z}^i \right] \,.
\label{C-qqbar}
\end{equation}

Inserting the short-distance coefficient (\ref{C-qqbar}) into
the factorization formula (\ref{dsig-fact}) and integrating over 
the phase space of the quarkonium, we get an expression 
for the inclusive cross section:
\begin{eqnarray}
\sum_X \sigma(q \bar q \to H + X)
&& 
\nonumber \\
&& \hspace{-1in} 
\;=\;  \delta(s - 4 m_c^2) \;
	{\pi^3 \alpha_s^2 \over 36 m_c^4} 
	\left[ \delta^{ji} - {\hat z}^j {\hat z}^i \right] 
	\langle \chi^\dagger \sigma^j T^a \psi \; 
		{\cal P}_H \;
		\psi^\dagger \sigma^i T^a \chi \rangle \,.
\label{sig-qqbar}
\end{eqnarray}
This is a term in the factorization formula for the inclusive cross section 
of any quarkonium state $H$, whether polarized or unpolarized. 
There are additional terms of order 
$\alpha_s^2$ from the process $gg \to c \bar c$.  All other parton processes
give terms with short-distance coefficients of order $\alpha_s^3$ or higher.

\section{NRQCD Matrix Elements}
\noindent
The long-distance factors in the NRQCD factorization formulas
are expressed as matrix elements of local four-fermion operators in
NRQCD.   Since long-distance effects in QCD are inherently nonperturbative,
the NRQCD matrix elements can only be calculated using nonperturbative
methods like lattice gauge theory.  There are effective lattice
prescriptions for calculating the matrix elements 
$\langle H | {\cal O}_{mn} | H \rangle$ that appear in 
quarkonium decay rates\cite{B-K-S}.
Unfortunately, these methods cannot be used to calculate directly
the production matrix elements $\langle {\cal O}^H_{mn} \rangle$.
The problem lies in implementing on the lattice
the projection defined by (\ref{P-H}). 
In the absence of nonperturbative calculations, the only alternative
is to treat the NRQCD matrix elements as phenomenological parameters
to be determined by experiment.  

\subsection{Model-independent framework}
\noindent
The factorization formula (\ref{dsig-fact})
provides a  model-independent framework for analyzing quarkonium 
production.  In any reasonable model for the production of quarkonium
through short-distance parton processes, the inclusive cross section can 
be expressed in the factored form (\ref{dsig-fact}).  The model
can therefore be reduced to a set of 
assumptions about the  NRQCD matrix elements.
Until recent years, most calculations of quarkonium production 
were carried out using either the {\it color-singlet model}
or the {\it color-evaporation model}\cite{Schuler}.
In the color-singlet model, the quarkonium is assumed to be
simply a color-singlet $c \bar c$ pair in an appropriate 
angular-momentum state. Only one NRQCD matrix element
is assumed to be important, and it 
can be expressed in terms of the $c \bar c$ wavefunction, 
or one of its derivatives, evaluated at the origin.
In the color-evaporation model, the color and angular-momentum 
quantum numbers of the quarkonium are simply ignored.
The NRQCD matrix elements are assumed
to be calculable in perturbation theory up to an overall 
normalization constant that depends on the state $H$. 
This model implies that the  matrix elements 
scale like $v^{3+D}$, where $v$ is a small parameter
and $D$ is the number of 
covariant derivatives ${\bf D}$ in the operator ${\cal O}^H_{mn}$.

NRQCD predicts a much more intricate hierarchy among the matrix elements.
The matrix element $\langle {\cal O}^H_{mn} \rangle$ defined
in (\ref{O-H}) scales like $v^{3+D+E+2M}$, where $D$ is the number of 
covariant derivatives ${\bf D}$ that appear in the operator and 
$E$ and $M$ are the number of chromoelectric and chromomagnetic
transitions that are required for $c \bar c$ pairs in the states created
by $\psi^\dagger {\cal K}_m \chi$ and $\psi^\dagger {\cal K}_n \chi$
to reach the dominant Fock state of $H$.
These velocity-scaling rules determine the 
approximate magnitudes of NRQCD
matrix elements.  By keeping only those matrix elements that scale 
with the fewest powers of $v$, we can reduce their number
sufficiently that a phenomenological approach becomes tractable.
One should keep in mind, however, that the importance of  a term 
in the cross section is determined not only by the magnitude 
of the matrix element but also by the magnitude of its 
short-distance coefficient.

\subsection{Reducing the matrix elements}
\noindent
The matrix elements can be further simplified by using 
symmetries of NRQCD.  To illustrate the simplifications, we will use 
matrix elements of the $J/\psi(\lambda)$, where  
the polarization state is specified by the helicity $\lambda$.
 
\subsubsection{Rotational symmetry}
\noindent
Rotational symmetry is an exact symmetry of NRQCD.  
It implies, for example,  that the matrix element for $H = \psi(\lambda)$ in 
(\ref{sig-qqbar}) must be a linear combination of
the tensors $\delta_{ij}$, $U_{\lambda j} U^\dagger_{i \lambda}$,
and $U_{\lambda j} U^\dagger_{i \lambda}$, where $U_{i \lambda}$
is the unitary matrix that transforms vectors from the spherical basis to the
Cartesian basis. If that matrix element is summed over 
the polarizations of the $\psi$,
the only possible tensor is $\delta_{ij}$.  The matrix element
must therefore satisfy
\begin{equation}
\sum_\lambda \langle \chi^\dagger \sigma^j T^a \psi \; 
		{\cal P}_{\psi(\lambda)} 
		\psi^\dagger \sigma^i T^a \chi \rangle \;=\;
{\delta_{ij} \over 3} \langle \chi^\dagger \sigma^k T^a \psi \; 
		{\cal P}_\psi \;
		\psi^\dagger \sigma^k T^a \chi \rangle .
\label{rot-sym}
\end{equation}

\subsubsection{Heavy-quark spin symmetry}
\noindent
Heavy-quark spin symmetry is an approximate symmetry of NRQCD
that holds up to corrections of order $v^2$.  The symmetry follows from
the fact that the spin of the heavy quark
is conserved at leading order in $v^2$  in NRQCD.
It implies, for example,  that the matrix element for $H = \psi(\lambda)$ in 
(\ref{sig-qqbar}) must be proportional to
$U_{\lambda j} U^\dagger_{i \lambda}$:
\begin{equation}
\langle \chi^\dagger \sigma^j T^a \psi \; 
		{\cal P}_{\psi(\lambda)} 
		\psi^\dagger \sigma^i T^a \chi \rangle \;\approx\;
{U_{\lambda j} U^\dagger_{i \lambda} \over 3} 
\langle \chi^\dagger \sigma^k T^a \psi \; 
		{\cal P}_\psi \;
		\psi^\dagger \sigma^k T^a \chi \rangle .
\label{hqs-sym}
\end{equation}
Summing over helicities, we recover (\ref{rot-sym}).

\subsubsection{Vacuum-saturation approximation}
\noindent
The vacuum-saturation approximation can only be applied to 
specific color-singlet matrix elements.  
If the matrix element is expressed in the form (\ref{O-H}), the operators 
$\psi^\dagger {\cal K}_m \chi$ and $\psi^\dagger {\cal K}_n \chi$
must create pointlike $c \bar c$ pairs in 
the dominant Fock state of the quarkonium.
In the vacuum-saturation approximation, the projection operator
$P_H$ defined in (\ref{P-H}) is replaced by $ | H \rangle \langle H|$, 
which corresponds to keeping only the vacuum term 
in the sum over soft states $X_S$. 
This is a controlled approximation in NRQCD,
holding up to corrections that are of order $v^4$.
The vacuum saturation approximation can be illustrated  by
the following matrix element for $J/\psi$ production:
\begin{equation}
\langle 0 | \chi^\dagger \sigma^i \psi \; {\cal P}_{\psi(\lambda)} \;
	 \psi^\dagger \sigma^j \chi | 0 \rangle
\;\approx\;
\langle 0 | \chi^\dagger \sigma^i \psi | \psi(\lambda) \rangle  \; 
	\langle \psi(\lambda) | \psi^\dagger  \sigma^j \chi | 0 \rangle \,.
\label{vsa-prod}
\end{equation}
The vacuum-saturation approximation can also be used for 
decay matrix elements:
\begin{equation}
\langle \psi | \chi^\dagger \mbox{\boldmath{$\sigma$}} \psi \cdot
	 \psi^\dagger  \mbox{\boldmath{$\sigma$}} \chi | \psi \rangle
\;\approx\; \sum_\lambda 
\big| \langle \psi(\lambda) | \psi^\dagger \mbox{\boldmath{$\sigma$}} 
	\chi | 0 \rangle \big|^2 \,.
\label{vsa-decay}
\end{equation}
The vacuum-to-quarkonium matrix element
$\langle \psi | \psi^\dagger \mbox{\boldmath{$\sigma$}} \chi | 0 \rangle$ 
that appears
on the right sides of (\ref{vsa-prod}) and (\ref{vsa-decay})  is
proportional to the wavefunction at the origin.
This matrix element can be easily calculated using lattice simulations of NRQCD.
Thus the vacuum-saturation approximation provides a way to 
calculate certain production matrix elements on the lattice.

\section{Polarization Predictions}
\noindent
The NRQCD factorization formulas apply equally well to any 
quarkonium state $H$, including  a polarized state.
Since the short-distance coefficients are independent of $H$, the
dependence on $H$ enters only through the NRQCD matrix elements.
In particular, the dependence on the polarization
comes only from the matrix elements.  In many cases, it is
completely determined by the symmetries of NRQCD.

The color evaporation model predicts that quarkonium states 
are always produced unpolarized.  Both the color-singlet model and 
the NRQCD factorization approach give nontrivial predictions
for polarized quarkonium.
There have been many calculations of polarization effects in the 
color-singlet model.  In most cases, the resulting predictions are not 
particularly dramatic.  Below, we discuss
several examples for which the calculations have have 
been extended to include color-octet mechanisms 
predicted by the NRQCD factorization formalism.  
In one case, we find a very dramatic prediction.

\subsection{Spin alignment at the Tevatron}
\noindent
The NRQCD factorization framework has led to a dramatic change
in our understanding of the production of charmonium
at large transverse momentum in $p \bar p$ collisions\cite{B-F-Y}. 
As pointed out by Braaten and Yuan\cite{Braaten-Yuan} in 1993,
the cross section for $p \bar p \to \psi + X$
at sufficiently large transverse momentum $p_T$
is dominated by gluon fragmentation.  
It can be factored into 
the cross section for producing a gluon with large transverse momentum
and a fragmentation function: 
\begin{equation}
d \sigma (p \bar p \to \psi(P) + X) \;=\; 
\int_0^1 dz \; d \hat{\sigma} (p \bar p \to g(P/z) + X) \; 
	D_{g \to \psi}(z) \,.
\label{dsig-frag}
\end{equation}
The fragmentation function $D_{g \to \psi}(z)$ gives the probability
that the jet initiated by the gluon includes a $\psi$ carrying 
a fraction $z$ of the gluon momentum.  Using the NRQCD factorization
approach, the fragmentation function can be expressed in the form
\begin{equation}
D_{g \to \psi}(z) \;=\; 
\sum_{mn} d_{mn}(z) \langle {\cal O}_{mn}^H \rangle \,.
\label{frag-fact}
\end{equation}
The matrix element that is leading order in $v$ is
$|\langle \psi | \psi^\dagger \mbox{\boldmath{$\sigma$}} \chi | 0 \rangle|^2$,
which scales like $v^3$.  It has a short-distance coefficient of order 
$\alpha_s^3$.  Using this term in the fragmentation function (\ref{frag-fact}), 
the cross section predicted by (\ref{dsig-frag}) is about a factor of  
30 below recent data on prompt $\psi$ production at the Tevatron
from the CDF detector\cite{Sansoni}. 

In 1995, Braaten and Fleming\cite{Braaten-Fleming} suggested that 
the gluon fragmentation function for the $\psi$ 
might actually be dominated by a term that represents
a color-octet production mechanism.  The matrix element is
$\langle \chi^\dagger \sigma^k T^a \psi \; {\cal P}_\psi \;
	 \psi^\dagger   \sigma^k T^a \chi \rangle$,
which is of order $v^7$ 
and measures the probability of producing a $\psi$
from a pointlike $c \bar c$ pair in a color-singlet $^3S_1$ state.
The reason this matrix element might be important is that its
short-distance coefficient is of order $\alpha_s$.   
The enhancement from the two fewer powers of $\alpha_s$ 
can overcome the suppression by $v^4$.
The leading-order expression for this term in the fragmentation function is 
\begin{equation}
D_{g \to \psi}(z) \;=\; 
{\pi \alpha_s \over 96 m_c^4} \delta(1-z) \; 
\langle \chi^\dagger \sigma^k T^a \psi \; {\cal P}_\psi \;
	 \psi^\dagger   \sigma^k T^a \chi \rangle \,.
\label{D-psi}
\end{equation}
The $p_T$-dependence predicted by this  mechanism is in agreement 
with the CDF data.  The normalization depends on the unknown matrix element 
in (\ref{D-psi}).  The value of
the matrix element required to fit the CDF data is consistent with
suppression by a factor of $v^4$ relative to the corresponding color-singlet 
matrix element
$|\langle \psi | \psi^\dagger \mbox{\boldmath{$\sigma$}} \chi| 0 \rangle|^2$.

Cho and Wise\cite{Cho-Wise} pointed out in 1995 that this production
mechanism has dramatic implications for the polarization of the $\psi$.
At leading order in $\alpha_s$, the $\psi$'s produced by gluon
fragmentation will be 100\% transversely polarized.  
The radiative corrections were examined by Beneke and 
Rothstein\cite{Beneke-Rothstein}, who concluded that the spin alignment 
at large $p_T$ will remain greater than 90\%.
The dominant corrections to the spin alignment at the values of $p_T$
measured at the Tevatron come from
nonfragmentation contributions to the cross section\cite{Cho-Leibovich}.
At large $p_T$, they fall like $1/p_T^2$ relative to gluon fragmentation.  
Unfortunately, the calculations required to obtain the 
prediction for the spin alignment as a function of $p_T$ 
have not yet been carried out.

The predictions for the spin alignment of the $\psi'$ are identical 
to those for the  $\psi$.
The spin alignment for the $\psi'$ is easier to measure,
because, in the case of the $\psi$,
one has to take into account the effects of the radiative decays 
$\chi_{cJ} \to \psi \gamma$. It should be possible to make at least
a crude measurement of the spin alignment of the $\psi'$ 
from existing CDF data.

\subsection{Spin alignment in $Z^0$ decay}
\noindent
The color-singlet model predicts that the cross sections for 
producing charmonium in $Z^0$ decay are too small to be observed at LEP.
As pointed out by Cheung, Keung, and Yuan and by Cho\cite{C-K-Y}, 
the dominant contribution comes instead from a color-octet production 
mechanism involving the same matrix element that appears 
in the fragmentation function (\ref{D-psi}).  Using the value 
of this matrix element obtained by fitting the CDF data,
they found that the rate is almost an order of magnitude
larger than predicted by the color-singlet model 
and thus large enough to be observed.
The spin alignment of the $\psi$ in  $Z^0$ decay was also
calculated by  Cheung, Keung, and Yuan.
Unfortunately, the alignment is predicted to be small and is 
completely unobservable in the  present data sample from LEP.

\subsection{Spin alignment in $B$ decay}
\noindent
In the production of $\psi$ from $B$ decay,  color-singlet 
contributions are suppressed by a near cancellation between 
Wilson coefficients in the effective weak hamiltonian.
Color-octet production mechanisms 
are therefore important, in spite of the  $v^4$ suppression of the matrix 
elements\cite{K-L-S}.
The spin alignment of the $\psi$'s that are produced in the decay 
$B \to \psi + X$ were recently calculated by Fleming et 
al.\cite{F-H-M-N}.  It depends sensitively on the values 
of 3 independent color-octet matrix elements, including the one
that appears in (\ref{D-psi}).  A measurement of the spin alignment
would therefore place strong constraints on these matrix elements.

\section{Conclusions}
\noindent
The factorization formulas (\ref{dsig-fact}) and (\ref{dGam-fact})
provide a model-independent framework for analyzing heavy quarkonium 
production and annihilation rates.  All the short-distance factors can be 
calculated systematically using the threshold expansion method.
The long-distance factors are defined
in terms of  NRQCD matrix elements.
The decay matrix elements can be computed using lattice simulations 
of NRQCD, but most of the production matrix elements must be treated as
phenomenological parameters.  The relative magnitudes of the 
matrix elements are predicted by the velocity-scaling rules of NRQCD.
These magnitudes have a pattern that is very different from that
assumed in the color-singlet model or in the color-evaporation model.

The NRQCD factorization approach gives unambiguous predictions for the 
polarization of heavy quarkonium states.  These predictions are 
inescapable consequences of this framework. The polarization predictions 
from NRQCD factorization  can be dramatic.  An example is the 
spin alignment of prompt $\psi$ and $\psi'$ at the Tevatron,
which is predicted to be greater than 90\% at the largest values of $p_T$.
An experimental measurement of this spin alignment
would be a crucial test of the NRQCD factorization framework.

\nonumsection{Acknowledgements}
\noindent
This work was supported in part by the United States Department of Energy,
Division of High Energy Physics, under grant DE-FG02-91-ER40690.

\nonumsection{References}


\begin{thebibliography}{000}

\bibitem{B-B-L} 
G.T. Bodwin, E. Braaten, and G.P. Lepage, 
	{\bibit Phys. Rev.} {\bf D51}, 1125 (1995).

\bibitem{Braaten-Chen} 
E. Braaten and Y.-Q. Chen, OHSTPY-HEP-T-96-010 (hep-ph/9604237).

\bibitem{B-K-S} 
G.T. Bodwin, D.K. Sinclair, and S. Kim,  ANL-HEP-PR-96-28 (hep-lat/9605023).

\bibitem{Schuler} 
G.A. Shuler, CERN-TH.7170/94 (hep-ph/9403387),
to appear in {\bibit Physics Reports}.

\bibitem{B-F-Y}
E. Braaten, S. Fleming, and T.C. Yuan, OHSTPY-HEP-T-96-001 (hep-ph/9602374), 
to appear in {\bibit Annual Reviews of Nuclear and Particle Science}.

\bibitem{Braaten-Yuan} 
E. Braaten and T.C. Yuan, {\bibit Phys. Rev. Lett.}  {\bf 71}, 1673 (1993).

\bibitem{Sansoni}
A. Sansoni et al. (CDF Collaboration), 
FERMILAB-CONF-95/263-E, to appear in the 
Proceedings of the Sixth International Symposium on Heavy Flavor Physics.

\bibitem{Braaten-Fleming} 
E. Braaten and S. Fleming, {\bibit Phys. Rev. Lett.} {\bf 74}, 3327 (1995).

\bibitem{Cho-Wise} 
P. Cho and M. Wise, {\bibit Phys. Lett.} {\bf B346}, 129 (1995).

\bibitem{Beneke-Rothstein} 
M. Beneke and I.Z. Rothstein, {\bibit Phys. Lett.} {\bf B372}, 157 (1996).

\bibitem{Cho-Leibovich} 
P. Cho and A.K. Leibovich, {\bibit Phys. Rev.} {\bf D53}, 6203 (1996).

\bibitem{C-K-Y} 
K. Cheung, W.-Y. Keung, and T.C. Yuan, 
	{\bibit Phys. Rev. Lett.} {\bf 76}, 877 (1996);
P. Cho, {\bibit Phys. Lett.} {\bf B368}, 171 (1996).

\bibitem{K-L-S} 
P. Ko, J. Lee, and H.S. Song, {\bibit Phys. Rev.} {\bf D53}, 1409 (1996).

\bibitem{F-H-M-N} 
S. Fleming et al., MADPH-96-953 (hep-ph/9608413).

\end{thebibliography}
\end{document}